%% file: paper.tex
\documentclass[letterpaper,twocolumn,10pt]{article}
\usepackage{usenix,epsfig,endnotes}
\usepackage{opera}
\usepackage{hyperref}
\usepackage{pgf, tikz, graphicx}
\usepackage{mathtools}
\usepackage{subfig}
\usetikzlibrary{arrows, automata}
\usepackage{comment}
\usepackage{url}
\usepackage[numbers,sort]{natbib}
\usepackage[font={footnotesize}]{caption}

\expandafter\def\expandafter\UrlBreaks\expandafter{\UrlBreaks
  \do\a\do\b\do\c\do\d\do\e\do\f\do\g\do\h\do\i\do\j%
  \do\k\do\l\do\m\do\n\do\o\do\p\do\q\do\r\do\s\do\t%
  \do\u\do\v\do\w\do\x\do\y\do\z\do\A\do\B\do\C\do\D%
  \do\E\do\F\do\G\do\H\do\I\do\J\do\K\do\L\do\M\do\N%
  \do\O\do\P\do\Q\do\R\do\S\do\T\do\U\do\V\do\W\do\X%
  \do\Y\do\Z}

\newcommand{\eat}[1]{}

\usepackage{titlesec}
\titlespacing*{\section}{0pt}{0.3\baselineskip}{0.1\baselineskip}
\titlespacing*{\subsection}{0pt}{0.3\baselineskip}{0.1\baselineskip}

\begin{document}

\date{}
\title{\Large \bf FRAPpuccino: Fault-detection through Runtime Analysis of Provenance}
\author{
{\rm Xueyuan Han, Thomas Pasquier, Tanvi Ranjan, Mark Goldstein, Margo Seltzer}\\
Harvard University
} 

\maketitle
\thispagestyle{empty}

\subsection*{Abstract}
\input{000-abstract}

\section{Introduction}
\label{sec:introduction}
\input{new-introduction}

\section{Background}
\label{sec:background}
\input{020-background}

\section{The FRAPpuccino Framework}
\label{sec:framework}
\input{new-framework}

\subsection{Learning Stage}
\label{sec:learning}
\input{new-learning}

\subsection{Detection Stage}
\label{sec:detecting}
\input{new-detecting}

\subsection{Revision Stage}
\label{sec:revising}
\input{033-revising}

\section{Implementation and Results}
\label{sec:implementation}
\input{new-implementation}

\section{Related Work}
\label{sec:related}
\input{new-related}

\section{Conclusion}
\label{sec:conclusion}
\input{070-conclusion}

\section*{Discussion Topics}
\input{080-discussion}

\section*{Acknowledgements}
\label{sec:ack}
\input{ack}

{\footnotesize \bibliographystyle{acm}
\bibliography{biblio}}

\appendix

\section{Availability}
\label{sec:availability}
\input{999-availability}

\end{document}

%% file: 000-abstract.tex
We present FRAPpuccino (or FRAP), a provenance-based fault detection mechanism for
Platform as a Service (PaaS) users, who run many instances of an application on a large cluster of machines.
FRAP models, records, and analyzes the behavior of an application and its impact on the system as a directed acyclic provenance graph.
It assumes that most instances behave normally and uses their
behavior to construct a model of \emph{legitimate behavior}.
Given a model of legitimate behavior, FRAP uses
a \textit{dynamic sliding window} algorithm to compare a new instance's
execution to that of the model.
Any instance that does not conform to the model is identified as an anomaly.
We present the FRAP prototype and experimental results showing that it can
accurately detect application anomalies.

%% file: new-introduction.tex
Platform as a service (PaaS) clouds have become increasingly popular for their efficient use of computational resources, 
providing users with an abstracted environment on which to easily deploy customized applications.
Various market research companies estimate the growth of the PaaS market at about 30\% annually for the next few years~\cite{zionglobal, digimonica}.
However, PaaS cloud applications face two major challenges: 
1) As an increasing number of businesses, enterprises, and organizations adopt cloud computing, 
cloud applications inevitably become a major target of cyber-attacks. 
For example, a recent DDoS attack against a top security blogger delivered by hijacked botnet was so aggressive that Akamai had to cancel the account~\cite{ddosattack}. 
According to the RightScale 2017 State of the Cloud Report~\cite{rightscale}, cloud security remains one of the top 5 challenges among cloud users; 
2) PaaS clouds make it possible to build large-scale applications that can serve millions of users.
A run-time fault introduced in an application can potentially render it useless.
For example, a simple bug can repeatedly crash a server,
making the service appear unavailable (~\autoref{sec:implementation}).

We introduce \textbf{FRAP}puccino (\textbf{F}ault-detection through \textbf{R}untime \textbf{A}nalysis of \textbf{P}rovenance or FRAP for short), a fault/intrusion detection framework.
FRAP detects anomalous behavior of cloud applications
by using runtime provenance data to model correct program execution and detecting
deviations from the model to identify potentially malicious behavior.
Provenance describes system behavior as a labelled directed acyclic graph (DAG) representing interactions between system-level \emph{entities} (\eg file, sockets, pipes), \emph{activities} (\ie processes) and \emph{agents} (\ie users, groups).
Provenance data can be abundant, so
FRAP implements a streaming algorithm, using a \textit{dynamic sliding window}, to
avoid storing this data.
From each running instance, the algorithm extracts a feature vector, which is
a projection of the graph as a point into an $n$-dimensional space.
We assume that most instances exhibit legitimate behavior
so that clustering on these features will clearly divide the instances into
good and bad sets.
Thus, our application model includes two parts: the extracted features and the parameters of
the clusters.
Once FRAP has constructed such a model, it monitors program
executions, 
extracts features from them, and reports any instances whose features deviate significantly
from \emph{good} behavior.

Unlike most behavioral-based intrusion/fault detection systems~\cite{modi2013survey}
that rely on system-call usage~\cite{forrest1996sense, somayaji2002operating, sharif2007understanding, feng2004formalizing, xu2016sharper} to profile legitimate application behavior,
FRAP uses provenance data that provides a more comprehensive view of program activities,
including their effects on the underlying system.
Prior research has shown that understanding the context of a program's execution, 
which the provenance DAG provides,
leads to greater accuracy in detecting program anomalies~\cite{sekar2001fast, feng2003anomaly, xu2016sharper}.
Moreover, the provenance records provide data that can be analyzed to assist in root cause analysis, ideally providing actionable information.
Our use of end-to-end provenance capture to detect intrusions or faults differs from other end-to-end tracing approaches in two major ways:
1) by using runtime graphical and statistical analysis on provenance DAGs of normal instances of an application, 
FRAP requires no application instrumentation or annotation,
while systems such as Pip ~\cite{reynolds2006pip} need developer-provided 
specifications of expected behavior;
2) FRAP analyzes interactions between potentially all executing applications and the system
as naturally presented by provenance DAGs,
while systems such as Magpie~\cite{barham2004using} and SpectroScope~\cite{sambasivan2011diagnosing} use event logs, which represent a carefully curated subset of system activity, which may or may not capture the key actions.
Our goal is to show an alternative approach and evidence of its efficacy in
tackling a long-standing problem of intrusion/fault detection.

The contributions of this work are:
1) a novel approach that combines provenance and graphical and statistical analysis to model the behavior of cloud applications;
2) a \textit{dynamic sliding window} algorithm that allows efficient processing of large provenance data to achieve online detection; 
and 3) an implementation of our framework with demonstrated accuracy.

%% file: 020-background.tex
We build FRAP using two existing open-source tools:
1) CamFlow~\cite{pasquier2016information, thomas_pasquier_2017_571427}, a state-of-the-art provenance capture system; and 
2) GraphChi~\cite{kyrola2012graphchi, graphchigithub}, a highly-efficient graph processing framework.
However, the concepts are not tied to either implementation.

\noindgras{CamFlow~\cite{pasquier2016information}:} 
Provenance records the chronology of ownership, change, and movement of an object or a resource. 
We use provenance data to understand the interactions between the monitored application, other applications, and the underlying operating system.
There are many provenance capture systems available, including PASS~\cite{muniswamy2006provenance}, Hi-Fi~\cite{pohly2012hi}, Linux Provenance Module~\cite{bates2015trustworthy}, and CamFlow~\cite{pasquier2016information}. 
We chose to use CamFlow, because it tracks multiple applications, their
interactions with the system, and their interactions with each other.
Moreover, it both limits the amount of information captured (\ie you can specify
which applications to trace) and ensures completeness by propagating capture
to any programs that a traced application invokes.
The capture is built upon Linux Security Module (LSM) hooks that provide completeness guarantees ~\cite{edwards2002runtime, ganapathy2005automatic, jaeger2004consistency}.
CamFlow also provides a facility to conveniently stream the provenance data captured through messaging middleware such as MQTT\cite{mqttorg}, RabbitMQ\cite{rabbitmq}, or Apache Flume\cite{flumeapache}.

\noindgras{GraphChi~\cite{kyrola2012graphchi}:} 
Provenance data is naturally represented as a DAG in which each node represents an entity, an activity, or an agent, 
and each directed edge represents an interaction between two nodes. 
For example, a file \textit{was generated by} a process, or a process \textit{used} a packet. 
Our framework uses type information in a provenance record as labels to construct a labeled DAG. 
DAGs can be efficiently processed by graph processing engines. 
We chose GraphChi, a vertex-centric graph processing model, to generate program models and to detect anomalies. 
GraphChi uses a \textit{parallel sliding window} method to achieve efficient computation of vertices. 




%% file: new-framework.tex
The FRAP	 framework consists of three stages:
1) the \textit{learning stage} determines the size of the \textit{dynamic sliding window}
and creates a model of correct program behavior;
2) the \textit{detection stage} periodically compares instances of a program's execution with the model, notifying the user upon detection of
unusual program behavior;
and 3) the \textit{revision stage} improves the model by incorporating additional information when we encounter a false positive.

FRAP starts with the learning stage, iterates through the detection stage until it needs to revise the model, and transitions back to the detection stage after the revision.

%% file: new-learning.tex
The learning stage analyzes the provenance DAG of each instance of a program, 
creating a model to describe its legitimate behavior.
Provenance data can grow infinitely large, making it impossible to analyze them as a whole.
However, past research~\cite{forrest1996sense, hofmeyr1998intrusion} has shown that 
a program usually has a limited set of interactions with the system (\eg writing to a file, sending a packet), repeating them in different orders as it executes.
Therefore, we claim that one can learn most of a program's behavior from a subset of its provenance data.
We begin with an overview of our approach, followed by a more detailed discussion of each step.
FRAP uses a simple but effective algorithm to determine the number of consecutive provenance records it needs to examine to create a program model, which is
the \textit{dynamic sliding window} size (used in the detection stage).
Using this subset of records, FRAP transforms the provenance DAG into a multidimensional numerical feature vector.
We construct this feature vector in three steps:
First, we run a label propagation algorithm that constructs a label for each
node, representing the structure of the graph around the node.
Second, we count the number of instances of each unique label.
Third, we construct a feature vector consisting of all the label counts.
A feature vector is therefore the result of dimensionality reduction, abstracting a program's behavior into numerical values.
Finally, FRAP clusters the feature vectors with the goal of grouping all
well-behaved instances together and leaving badly behaved instances in
different clusters.
Thus, our model consists of the feature vectors of well behaved
instances and the parameters (centroid and radius) of the clusters in which
they reside.

\setlength{\textfloatsep}{8pt plus 1.0pt minus 2.0pt}

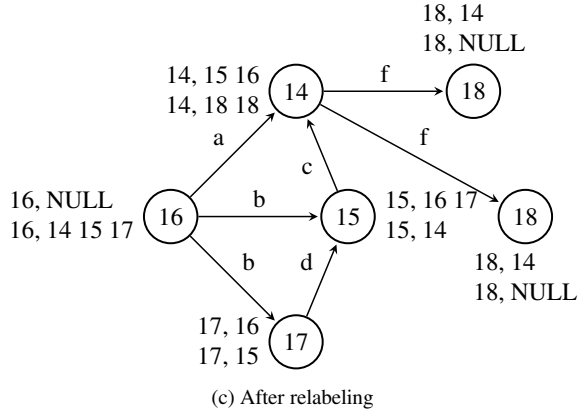
\begin{figure}[ht!]
\begin{center}
\subfloat[Before relabeling] {
\resizebox{\columnwidth}{!} {
\begin{tikzpicture}[
            > = stealth, 
            shorten > = 1pt, 
            auto,
            node distance = 2.5cm, 
            semithick 
        ]

        \tikzstyle{every state}=[
            draw = black,
            thick,
            fill = white,
            minimum size = 4mm
        ]

        \node[state] (s) [label={[align=left]left: 2, NULL\\2, 0 a 1 b 2 b}] {$2$};
        \node[state] (v1) [above right of=s] [label={[align=left]above: 0, 1 c 2 a\\0, 3 f}]{$0$};
        \node[state] (v2) [right of=s] [label={[align=left]right: 1, 2 b 2 d\\1, 0 c}]{$1$};
        \node[state] (v3) [below right of=s] [label={[align=left]left: 2, 2 b\\2, 1 d}]{$2$};
        \node[state] (v4) [right of=v1] [label={[align=left]right: 3, 0 f\\3, NULL}]{$3$};
        \node[state] (t) [right of=v2] [label={[align=left]below: 3, 0 f\\3, NULL}]{$3$};

        \path[->] (s) edge node {a} (v1);
        \path[->] (s) edge node {b} (v2);
        \path[->] (s) edge node {b} (v3);
        \path[->] (v2) edge node {c} (v1);
        \path[->] (v3) edge node {d} (v2);
        \path[->] (v1) edge node {f} (v4);
        \path[->] (v1) edge node {f} (t);
        
\end{tikzpicture}
}
\label{fig:before}
}
\hfill
\subfloat[Populating Relabeling map] {
\resizebox{\columnwidth}{!} {
\begin{tabular}{ l l l}
\multicolumn{3}{c} {Relabeling Map} \\
\hline
 \color{red} 0, 1 c 2 a $\rightarrow$ 4 & \color{blue} 0, 3 f $\rightarrow$ 5 & 4, 5 $\rightarrow$ 14\\ 
 \color{red} 1, 2 b 2 d  $\rightarrow$ 7 & \color{blue} 1, 0 c $\rightarrow$ 6 & 7, 6 $\rightarrow$ 15\\  
 \color{red} 2, NULL $\rightarrow$ 8 & \color{blue} 2, 0 a 1 b 2 b $\rightarrow$ 9 & 8, 9 $\rightarrow$ 16\\
 \color{red} 2, 2 b $\rightarrow$ 11 & \color{blue} 2, 1 d $\rightarrow$ 10 & 11, 10 $\rightarrow$ 17\\
 \color{red} 3, 0 f $\rightarrow$ 13 &  \color{blue} 3, NULL $\rightarrow$ 12 & 13, 12 $\rightarrow$ 18 
\end{tabular}
}
\label{table:map}
}
\hfill
\subfloat[After relabeling] {
\resizebox{\columnwidth}{!} {
\begin{tikzpicture}[
            > = stealth, 
            shorten > = 1pt, 
            auto,
            node distance = 2.5cm, 
            semithick 
        ]

        \tikzstyle{every state}=[
            draw = black,
            thick,
            fill = white,
            minimum size = 4mm
        ]

        \node[state] (s) [label={[align=left]left: 16, NULL\\16, 14 15 17}]{$16$};
        \node[state] (v1) [above right of=s] [label={[align=left]left: 14, 15 16\\14, 18 18}]{$14$};
        \node[state] (v2) [right of=s] [label={[align=left]right: 15, 16 17\\15, 14}]{$15$};
        \node[state] (v3) [below right of=s] [label={[align=left]left: 17, 16\\17, 15}]{$17$};
        \node[state] (v4) [right of=v1] [label={[align=left]above: 18, 14\\18, NULL}]{$18$};
        \node[state] (t) [right of=v2] [label={[align=left]below: 18, 14\\18, NULL}]{$18$};

        \path[->] (s) edge node {a} (v1);
        \path[->] (s) edge node {b} (v2);
        \path[->] (s) edge node {b} (v3);
        \path[->] (v2) edge node {c} (v1);
        \path[->] (v3) edge node {d} (v2);
        \path[->] (v1) edge node {f} (v4);
        \path[->] (v1) edge node {f} (t);
        
\end{tikzpicture}
\label{fig:after}
}
}
\caption{(a) Before the first iteration, we label each node with its own label, its neighboring nodes' labels, and its incident edges' labels. (b) Each node inserts its sorted in-edge neighbor label list and its sorted out-edge neighbor label list into the relabeling map, and gets a new label based on these two label lists. In-edge node relabeling is shown in red, out-edge node relabeling is shown in blue, and final relabeling is shown in black. (c) After relabeling, nodes have both new identities and new labels while edge labels are unchanged.}
\vspace{-1.5em}
\label{fig:oneIteration}
\end{center}
\end{figure}

\noindgras{Determining the Dynamic Sliding Window Size: }
The goal of finding a dynamically sized window is to minimize the number
of provenance records needed to characterize an application.
We determine the size during capture by maintaining two counts.
The first counts every edge examined until we declare a window size.
We calculate the second by the following process:
We examine each edge and assign it a triple 
consisting of the original edge type and the types of each vertex.
We count the number of edges processed until we encounter a triple
we have never seen before, at which point we reset the count to 0.
If the first counter reaches an implementation-defined threshold
(\autoref{sec:implementation}), or
if the second counter reaches a user-defined threshold,
we set the dynamic window size to be the value of the first counter.

\noindgras{Generating a Program Model: }
To generate a model, 
FRAP computes a feature vector for each instance, clusters those vectors, and discards the vectors in isolated clusters.
FRAP generates a feature vector, using the following iterative algorithm:
relabel each vertex in the DAG using the current vertex label and the labels of its incoming and outgoing neighbors.
During the first iteration, we also incorporate the labels of a vertex's incident edges (~\autoref{fig:oneIteration}), but need not include these in
later iterations, because that information is already encoded in the labels of the neighboring vertices.
After each iteration, new labels encode longer sequences of interactions between the program and the system.
FRAP generates a vector containing counts of all seen labels, including the ones from previous iterations, and then clusters the vectors from all instances.
We empirically determine that four iterations produces the best results.
To cluster, FRAP uses symmetric Kullback-Leibler divergence~\cite{kullback1951information} or \textit{Kullback-Leibler Distance} (KLD) with \textit{back-off probability} ~\cite{mori1997spoken} as the distance metric to measure the similarity between two feature vectors.
KLD has been used, for example, in statistical language modeling ~\cite{dagan1999similarity} and text categorization~\cite{bigi2003using}. 
We use two applications of K-means clustering: first we cluster on distances between feature vectors, which helps us select $K$ for the second K-means clustering, which computes the actual model.

We assume that clusters containing many vectors represent legitimate behavior
and want to discard feature vectors in clusters isolated from these good clusters.
Wagstaff et al. ~\cite{wagstaff2001constrained} have shown that cluster accuracy improves when additional information is available to the problem domain.
We use our assumption that most instances are well-behaved as this
additional information.
Specifically, we hypothesize that there exists an observable difference between \emph{inter-cluster} and \emph{intra-cluster} distances.
First  we set $K$ equal to the total number of instances we are analyzing and run K-means clustering on the \emph{pairwise KL distance between each pair of instances}.
We then set $K$ to the number of populated clusters and run a second K-means clustering on the feature vectors themselves.
This produces our model consisting of the set of feature vectors in clusters containing more than one instance and the parameters of those clusters (e.g., centroid and radius).

%% file: new-detecting.tex
FRAP monitors an instance by taking its provenance data from a window of execution, generating a feature vector as in \autoref{sec:learning}, and checking whether this vector fits into any clusters by comparing the distance between the feature vector and the centroids of the clusters. An instance is considered abnormal if it does not fit into any of the model's clusters.

\begin{figure}
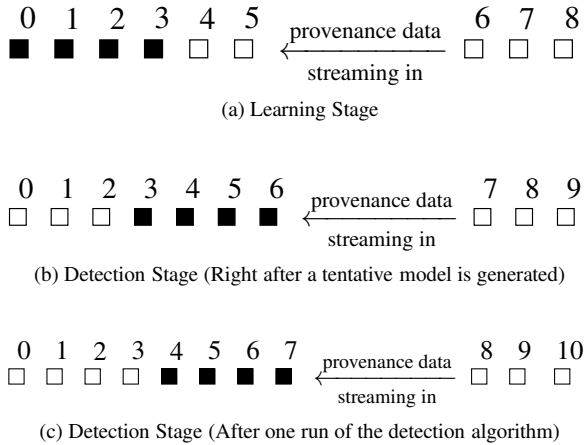

\begin{center}
\textbf{Example Detection Algorithm (Window Size $=$ 4)}
\subfloat[Learning Stage] {
\resizebox{\columnwidth}{!} {
\tikz{\path[draw=black,fill=black] (0,0) rectangle (0.2cm,0.2cm) node[align=left, above] {$0$};}
\tikz{\path[draw=black,fill=black] (0,0) rectangle (0.2cm,0.2cm) node[align=left, above] {$1$};} 
\tikz{\path[draw=black,fill=black] (0,0) rectangle (0.2cm,0.2cm) node[align=left, above] {\\$2$};} 
\tikz{\path[draw=black,fill=black] (0,0) rectangle (0.2cm,0.2cm) node[align=left, above] {\\$3$};} 
\tikz{\path[draw=black,fill=white] (0,0) rectangle (0.2cm,0.2cm) node[align=left, above] {\\$4$};}  
\tikz{\path[draw=black,fill=white] (0,0) rectangle (0.2cm,0.2cm) node[align=left, above] {\\$5$};} 
\text{$\xleftarrow[\text{streaming in}]{\text{provenance data}}$} \ 
\tikz{\path[draw=black,fill=white] (0,0) rectangle (0.2cm,0.2cm) node[align=left, above] {\\$6$};}
\tikz{\path[draw=black,fill=white] (0,0) rectangle (0.2cm,0.2cm) node[align=left, above] {\\$7$};}
\tikz{\path[draw=black,fill=white] (0,0) rectangle (0.2cm,0.2cm) node[align=left, above] {\\$8$};}
}
\label{fig:learning}
}
\hfill
\subfloat[Detection Stage (Right after a tentative model is generated)] {
\resizebox{\columnwidth}{!} {
\tikz{\path[draw=black,fill=white] (0,0) rectangle (0.2cm,0.2cm) node[align=left, above] {$0$};}
\tikz{\path[draw=black,fill=white] (0,0) rectangle (0.2cm,0.2cm) node[align=left, above] {$1$};} 
\tikz{\path[draw=black,fill=white] (0,0) rectangle (0.2cm,0.2cm) node[align=left, above] {\\$2$};} 
\tikz{\path[draw=black,fill=black] (0,0) rectangle (0.2cm,0.2cm) node[align=left, above] {\\$3$};} 
\tikz{\path[draw=black,fill=black] (0,0) rectangle (0.2cm,0.2cm) node[align=left, above] {\\$4$};}  
\tikz{\path[draw=black,fill=black] (0,0) rectangle (0.2cm,0.2cm) node[align=left, above] {\\$5$};} 
\tikz{\path[draw=black,fill=black] (0,0) rectangle (0.2cm,0.2cm) node[align=left, above] {\\$6$};}
\text{$\xleftarrow[\text{streaming in}]{\text{provenance data}}$} \ 
\tikz{\path[draw=black,fill=white] (0,0) rectangle (0.2cm,0.2cm) node[align=left, above] {\\$7$};}
\tikz{\path[draw=black,fill=white] (0,0) rectangle (0.2cm,0.2cm) node[align=left, above] {\\$8$};}
\tikz{\path[draw=black,fill=white] (0,0) rectangle (0.2cm,0.2cm) node[align=left, above] {\\$9$};}
}
\label{fig:detection1}
}
\hfill
\subfloat[Detection Stage (After one run of the detection algorithm)] {
\resizebox{\columnwidth}{!} {
\tikz{\path[draw=black,fill=white] (0,0) rectangle (0.2cm,0.2cm) node[align=left, above] {$0$};}
\tikz{\path[draw=black,fill=white] (0,0) rectangle (0.2cm,0.2cm) node[align=left, above] {$1$};} 
\tikz{\path[draw=black,fill=white] (0,0) rectangle (0.2cm,0.2cm) node[align=left, above] {\\$2$};} 
\tikz{\path[draw=black,fill=white] (0,0) rectangle (0.2cm,0.2cm) node[align=left, above] {\\$3$};} 
\tikz{\path[draw=black,fill=black] (0,0) rectangle (0.2cm,0.2cm) node[align=left, above] {\\$4$};}  
\tikz{\path[draw=black,fill=black] (0,0) rectangle (0.2cm,0.2cm) node[align=left, above] {\\$5$};} 
\tikz{\path[draw=black,fill=black] (0,0) rectangle (0.2cm,0.2cm) node[align=left, above] {\\$6$};}
\tikz{\path[draw=black,fill=black] (0,0) rectangle (0.2cm,0.2cm) node[align=left, above] {\\$7$};}
\text{$\xleftarrow[\text{streaming in}]{\text{provenance data}}$} \ 
\tikz{\path[draw=black,fill=white] (0,0) rectangle (0.2cm,0.2cm) node[align=left, above] {\\$8$};}
\tikz{\path[draw=black,fill=white] (0,0) rectangle (0.2cm,0.2cm) node[align=left, above] {\\$9$};}
\tikz{\path[draw=black,fill=white] (0,0) rectangle (0.2cm,0.2cm) node[align=left, above] {\\$10$};}
}
\label{fig:detection2}
}
\end{center}
\vspace{-1.5em}
\caption{Black rectangles represent the provenance records in the window. White ones to the left of the arrow are captured but not yet processed, and to the right will be generated in the future.
}
\vspace{-0.5em}
\label{fig:window}
\end{figure}

\noindgras{Dynamic Sliding Window: } FRAP uses a \textit{dynamic sliding window} approach to continuously monitor an instance while it runs uninterrupted. 
FRAP only stores and analyzes the provenance DAG within this window.
Once it determines that the DAG is part of a normal program execution, FRAP can safely discard the data.
The size of the window is determined in the learning stage (\autoref{sec:learning}).
As shown in \autoref{fig:window}, records $0, 1, 2, 3$ in \ref{fig:learning} are used to generate a model while new provenance data (records $4$ and $5$) are streaming in. 
The window slides each time FRAP finishes running the detection algorithm (described in the next paragraph), to include new records for the next round of analysis (\autoref{fig:detection1} and \autoref{fig:detection2}). 
Since the window slides incrementally, each record appears in \emph{many} different windows, maintaining a holistic view of program execution.
New records may contain vertices related to ones processed in previous windows.
This overlapping preserves a vertex's $n$-hop neighbors in which $n$ becomes larger as the overlap increases.
This means that a larger number of iterations becomes more meaningful,
since longer sequences of program-system interactions can be preserved.

\noindgras{Detection Algorithm: }  Given a feature vector of a window of execution, FRAP uses the same distance metrics (\ie KLD) to see if the instance lies in an existing cluster (\ie the distance between it and the centroid is smaller than the cluster's radius).
If it does not lie in an existing cluster, FRAP re-runs (the second) K-means clustering, and only then reports the instance as an anomaly if it still does not lie within a cluster of legitimate executions.
Since the detection algorithm is entirely local between an instance and the model, the computation is parallelizable and scalable.

%% file: 033-revising.tex
FRAP identifies and collects false-positive instances and re-clusters, including them in the model.
This reclustering is fast, which is important, because
instances are running concurrently.
A fast transition from the revision stage to the detection stage ensures maximal overlap between the windows before and after revision (\autoref{sec:detecting}).

Our current false-positive identification is somewhat crude.
We assume that most instances are correct and revise the model only when detecting a large number of abnormal instances. 
However, this assumption does not always hold. 
For example, a corrupted database may cause many client requests to fail. FRAP should not learn such behavior and include it in the model.
Another problematic scenario is when a program has normal behavior ($A$, $B$, $C$), and the model contains only one behavior ($A$). 
When many instances are displaying behavior $B$ and one instance is displaying $C$, then only $B$ will be included in the model while that one instance will be considered unusual. Developing a more principled approach to this problem is left for future work.

%% file: new-implementation.tex
We describe our implementation~\cite{han_2017_571444}, experimental results and future work.

\noindgras{Implementation-specific Window Size: } When we constantly receive unique provenance records, our implementation-defined threshold determines the window size (\autoref{sec:learning}). GraphChi uses a parallel sliding window method to perform computation on the vertices inside the window in memory, write the results to disk, and move the window to load the next set of vertices to memory. To avoid high-latency I/O operations, the learning stage takes the provenance data of the maximum size of the memory for each instance and allows GraphChi to perform computation on all vertices in-memory. 

\noindgras{Program Models and Feature Vectors: } GraphChi performs vertex relabeling to generate a feature vector for each instance. Our implementation uses a global map to achieve consistent relabeling. We need to handle GraphChi's asynchronous model of computation, in which an update of a vertex label is immediately visible to its neighbors in the same iteration. This means if a vertex runs its computation \textit{after} one of its neighbors updates itself, it will take that neighbor's updated label, which should be used in the next iteration. We solve this problem by alternating the computation with an \textit{update phase} and a \textit{swap phase}. In the update phase (which consists of one iteration), each vertex computes its new label without broadcasting it to its neighbors. Therefore, within the same iteration, all vertices take their neighboring vertices' labels from the previous update phase. The next iteration is the swap phase, where all vertices broadcast their new labels to their neighbors so that in the next update phase, they can all read the latest labels.

\begin{table}[htb]
\begin{center}
\resizebox{\columnwidth}{!}{
\begin{tabular}{| l || p{2.5cm} | p{2.5cm} | p{2.5cm} |}
\hline
Metrics & Captured Bad \newline Instance \newline (Learning) & Captured Bad \newline Instance \newline (Detection) \\
\hline\hline
KLD & Yes & Yes \\
\hline
Hellinger & No & No \\
\hline
Euclidean & Yes & Yes \\
\hline
\end{tabular}
}
\caption{This experiment runs 10 clients sending requests to the server, one of which causes the server to behave abnormally during learning. The same bad behavior occurs again during detection.}
\vspace{-1.5em}
\label{table:results}
\end{center}
\end{table}

\noindgras{Preliminary Results: }
We conducted a number of experiments to see if FRAP is able to capture instances with unusual behavior in a pool of well-behaved instances. 
We used both Hellinger distance ~\cite{nikulin2001hellinger} and Euclidian distance, in addition to KLD, to see how different similarity metrics affect FRAP's performance. 
\autoref{table:results} shows the results of one of these experiments.
In this experiment, we set up a Ruby server in a simulated cloud environment.
The server handles requests from multiple clients, and causes an out-of-memory server crash\cite{rubyattack} -- a known system level Ruby vulnerability -- for certain URLs.

We see that all but Hellinger distance are able to identify the badly behaved Ruby instance.
Hellinger distance does not work well, because it always produces values
between $0$ and $1$, which make it difficult for
K-means clustering to create meaningful clusters. 
Hellinger consistently placed all instances in a single cluster.
After manual inspection, we also discovered that while Euclidean distance successfully identified the badly behaved instance,
it mistakenly considered two normal instances as abnormal during the learning stage (\ie false negatives), resulting in a slightly less accurate model than KLD.
\autoref{sec:availability} provides a reference to
the current prototype and other experimental datasets.

\noindgras{Future Work: } We plan to develop more sophisticated ways to identify false-positives as outlined in \autoref{sec:framework}.
We also want to further optimize our current implementation and identify situations in which more sophisticated learning algorithms will improve accuracy. 
One important area of optimization is our global relabeling map. 
Since GraphChi processes vertices in parallel, we need to maintain map consistency and avoid race conditions. 
Our current implementation 
is a single-point of contention and can easily be a performance bottleneck. 
We also have to garbage collect the map to keep its size manageable.
We also want to experiment other algorithms to improve our clustering. For example, Principle Component Analysis ~\cite{jolliffe2002principal} might help us further reduce dimensionality of feature vectors and discard misleading features.

Our current prototype does not have all the pieces integrated, so we used manual
intervention to achieve the end to end pipeline. In particular, we do not directly
stream the provenance to the analyzer. Instead, we capture the provenance and
then run the analyzer over the provenance stream.
Additionally, we have not yet integrated the revision stage.
These will both be available in the next release of the software.

%% file: new-related.tex
A number of systems use sequences of system calls to detect program/system anomalies. 
pH~\cite{somayaji2002operating} uses temporally proximate system call sequences to model the behavior of a program.
More recent systems have proposed more advanced analyses.
For example, MaMaDroid~\cite{mariconti2016mamadroid} uses static program analysis to obtain a program's call graph and dynamically builds a Markov-chain~\cite{kemeny1960finite} model of the graph during runtime. 
The feature vector of their program model consists of the probabilities of each state transition in the Markov chain. 
CMarkov ~\cite{xu2016sharper} includes calling context of system calls when performing static program analysis, which further refines their Markov model. 
FRAP uses provenance to build a program model without the help of static program analysis. 
Its relabelling mechanism concisely encodes system call sequences and their contexts.
Moreover, FRAP can detect anomalies in complex applications composed of multiple processes including distributed ones by capturing, aggregating, and analyzing provenance data from multiple machines. 
It is not restrained by per-process system call sequences. 
We leave comparing FRAP performance with that of other detection systems for future work.

Similar to FRAP, systems such as Magpie~\cite{barham2004using}, Pinpoint~\cite{chen2004path}, Pip~\cite{reynolds2006pip} and SpectroScope~\cite{sambasivan2011diagnosing} use end-to-end tracing to detect anomalies that could indicate bugs.
However, unlike FRAP, they all capture request flows within distributed systems and analyze event logs either on a per-event basis or on whole paths.
More importantly, although these systems are able to infer bugs, 
they are mainly designed to diagnose performance problems, not intrusions, in distributed systems.
Moreover, Pinpoint and Pip require manually annotating applications, an error-prone and significant burden on developers, 
while FRAP can analyze all applications as long as the underlying system captures provenance.

For systems that do not have detailed end-to-end tracing capabilities, some black-box diagnosis techniques have been proposed, \eg using message send/receive events to deploy black-box performance debugging~\cite{aguilera2003performance}.

Our graph analysis is related to work on graph kernels ~\cite{vishwanathan2010graph}. They are widely used in studying relationships between structured graphs ~\cite{vishwanathan2010graph}. In particular, our relabeling algorithm is based on the subtree Weisfeiler-Lehman graph kernel~\cite{shervashidze2011weisfeiler} and is a variation of the Weisfeiler-Lehman test of isomorphism~\cite{weisfeilerreduction}.

%% file: 070-conclusion.tex
We present a novel approach to detecting unusual behavior in programs running on PaaS clouds and demonstrate its usability via our implementation. We believe current advances in provenance capture systems open a new landscape for research in cloud computing and computer systems.

%% file: 080-discussion.tex
We assume that the provenance data we capture are trustworthy and that attackers cannot modify provenance data to mask the application's execution trail. 
What should we do if this assumption does not hold in practice? 
Our system requires mitigation techniques to guard against non-trustworthy provenance data or detect provenance data tampering as a different form of intrusion. 
This problem has been explored in the literature~\cite{braun2008securing, bates2015trustworthy}, 
and Zhou et al. ~\cite{zhou2011secure} presented a solution to secure network provenance. 
Can we simply ``plug-and-play'' those mechanisms in our system? 
How do we comprehensively secure various sources of provenance used in our system?

We propose that user involvement can help build a better model by allowing users to identify false-positives. 
What should we do to provide users with \textit{meaningful} provenance information to assist their judgement? 
One possible solution is to apply differential provenance~\cite{chen2015differential} to explain the sources of anomalies. 
However, differential provenance has only been applied to network provenance. 
How do we apply this technique to other domains? 

There are a variety of intrusion detection systems (IDS) for the cloud environment. 
Modi et al. \cite{modi2013survey}~categorized them into eight different techniques, identifying both their strengths and weaknesses.
FRAP is a behavioral-based detection system, but unlike other systems in this category, it uses provenance to model the behavior of an application. 
From what aspects does FRAP work better than other behavioral-based detection systems and than other cloud IDS's at large? 
What kinds of intrusions are intrinsically hard for FRAP to detect but easy for other IDS's?

%% file: ack.tex
\noindent We thank Daniel Margo for inspiring discussions of graph processing frameworks. This work was supported by the US National Science Foundation under grant SSI-1450277 End-to-End Provenance.

%% file: 999-availability.tex
The work presented in this paper is open-source and available for download at
\begin{center}
\url{https://github.com/michael-hahn/frap}\\
\end{center}